\documentclass[10pt,letterpaper]{article}
\usepackage{graphics}
\usepackage{amsmath}
\usepackage{subfigure}
\usepackage[pdftex]{graphicx}
\usepackage{color}
\usepackage{hyperref}

\hypersetup{colorlinks,
            linkcolor=blue,
            anchorcolor=blue,
            citecolor=magenta
            }
\begin{document}

{\Large \textbf{Accurate determination of the quality factor and tunneling distance of axisymmetric resonators for biosensing applications}}\\

{\normalsize M. Imran Cheema, and Andrew G. Kirk}\\

 {\footnotesize ECE dept., McGill University, 3480 University Street, Montreal, Canada H3A 2A9}\\

{\scriptsize imran.cheema@mail.mcgill.ca} 
\section*{abstract}
Due to ultra high quality factor ($10^6-10^9$), axisymmetric optical  microcavities are popular platforms for biosensing applications. It has been recently demonstrated that a microcavity biosensor can track a biodetection event as a function of its quality factor by using phase shift cavity ring down spectroscopy (PS-CRDS). However, to achieve maximum sensitivity, it is necessary to optimize the microcavity parameters for a given sensing application. Here, we introduce an improved finite element model which allows us to determine the optimized geometry for the PS-CRDS sensor. The improved model not only provides fast and accurate determination of quality factors but also determines the tunneling distance of axisymmetric resonators.  The improved model is validated numerically, analytically, and experimentally.



\section{Introduction}
Microcavities are widely used in many applications such as biosensing, nonlinear processes, lasers and optomechanics. Most previously published work with microcavity biosensors has employed measurement of the change in resonant frequency of the sensor as a function of a biodetection event \cite{Vollmer2012, Fan_2008} with only a few instances of using the quality factor as a sensing metric \cite{Ilchenko_P2002,Ilchenko_2002,Armani_2006}. It has been shown that the quality factor of microcavities can be measured in the time domain using phase shift shift cavity ring down spectroscopy (PS-CRDS)\cite{Barnes_2008}. This approach has recently been applied to biosensing applications \cite{kirk2012}. In order to extend this work to ultra sensitive biosensing applications (e.g. determination of binding kinetics, detection of $pM-fM$ concentrations, and proteins, viruses etc ), it is important to design a cavity with the optimum parameters for high sensitivity. One of the goals of the current work, is to rapidly estimate the quality factors of the microcavities with high accuracy to address  these applications.

Whispering gallery mode (WGM) radiation occurs due to tunneling of photons through a potential barrier from the microcavity to a allowed wave propagation zone (see Fig. \ref {fig:potfield}). The tunneling phenomenon has been utilized in various photonic devices such as  photonic band gaps \cite{Spielmann1994},  super lattices \cite{Pedro2000} , and asymmetric cavities \cite{Podolskiy2005}. Recently Tomes et. al. \cite{Carmon2009} have performed remarkable experiments to image the tunneling process of axisymmetric microcavities. These experiments indicate that the tunneling phenomenon can be utilized for potential applications such as biosensing.  However, to design the appropriate devices for a particular biosensing application, it is again necessary to accurately calculate the tunneling distance of axisymmetric microcavities. Another goal of our work is to address this issue.

In any modeling technique, reflections from boundaries of the computation domain are induced due to radiation produced by the whispering gallery modes (WGM) of a microcavity. For an accurate electromagnetic model, absorbing boundary conditions or perfectly matched layers (PML) are required  to reduce these unwanted reflections. PML act as artificial boundaries that truncate the computation domain of open region scattering problems in the finite element method. There have been previous attempts to develop finite element models (FEM) of axisymmetric cavities which incorporate the PML. In 2005, Chinellato et al.  \cite{Chinellato2005} introduced a FEM model which was implemented in MATLAB. Whilst their model was capable of simulating the behavior of very small resonators ($< 3 \mu m$) it was not adequate for resonators possessing the dimensions typically used in experiments. In 2009, Karl et al. \cite{Karl2009} developed a 3D FEM model in JCMsuite for studying a micro-pillar cavity.  However, this model does not consider the suppression of false solutions, a well known problem in finite element formulations \cite{Suzuki1985, Osegueda94}. Moreover, the quality factor of the modes was estimated by fitting the Lorentzian peak to the calculated spectrum of the cavity, thus introducing an extra approximation.

In 2007, Oxborrow \cite{Oxborrow2007} developed a FEM for open axisymmetric resonators in COMSOL without invoking any transverse mode approximation to Maxwell's equations, representing an advance on previous work \cite{Spillane2004, Soltani2010}.  In his work, he showed that the model could simulate resonators of arbitrary cross section in optical and microwave regimes, thus removing the size limitation in \cite{Chinellato2005}. Another difference from \cite{Chinellato2005} is in terms of suppression of false modes; Oxborrow used a simple penalty term in his master equation whereas Chinellato et al. used  N\'{e}d\'{e}lec edge and modified Lagrange nodal element functions to avoid the spurious modes. With Oxborrow's formulation in COMSOL, 3D rotationally symmetric problems are reduced to 2D and are solvable in seconds which is vast improvement over \cite{Chinellato2005, Karl2009}. (For further discussion and further comparison to other works see references \cite {Oxborrow2007, Oxborrow2012}.)

The advantages of Oxborrow's model have made it a popular choice among researchers and it has been widely used in numerous research works e.g.\cite{{Oxborrow2012a},{Armani2012},{Vahala2011},{Gong2010}}. However in Oxborrow's model no PML was implemented and as a result the WGM quality factor could not be determined accurately. The quality factor due to the WGM radiation was estimated by placing a bound on its minimum and maximum possible values. These maximum and minimum values were determined by executing the model multiple times with different boundary conditions. Moreover, the model lacks the capability to estimate the quality factor of multiple modes simultaneously.

In order to provide accurate determination of the WGM quality factor, we have improved Oxborrow's model by modifying its master equation and implementing the PML along the boundaries of the computation domain. In the present work, geared towards sensing applications, we expand and refine the model presented in our earlier work \cite{Kirk2010}. Other researchers \cite{Goddard2010} have subsequently reported the inclusion of PML in Oxborrow's model for determining resonant frequencies and corresponding mode profiles of microring resonators but mathematical details have not been provided.

Our modified model does not have any of the drawbacks of Oxborrow's model. Moreover, we have computed the quality factors of all the modes without using any fitting algorithms as opposed to the approach taken in \cite{Karl2009}. Furthermore, with the modified Oxborrow's method, tunneling distances can also be accurately extracted for microcavities of various shapes. In the present work, a simple expression for computing tunneling distance of microtoroidal cavities is also provided.

In our model, we treat the PML as an anisotropic absorber and implement it in the cylindrical coordinate system. Our model is applicable to any axisymmetric resonator geometry but due to the availability of analytical expressions for spherical resonators, we have validated the model by determining the quality factors and tunneling distances of a silica microsphere in air. We have found that our simulation results are in excellent agreement with the analytical results. We also apply our model to microtoroidal cavities immersed in liquid and show that our results are consistent with those obtained by experiments. The model is then used to determine the optimum parameters of a microtoroidal cavity sensor based upon phase shift cavity ring down spectroscopy.

The paper is organized as follows. In section \ref {sec:math}, an improved FEM model is introduced by incorporating the PML. The optimal parameters of the PML are discussed in section \ref{sec:PML}. The analytical expressions of the quality factor and the tunneling distance of microspheres are presented in section \ref{sec:Analytical expressions of the spherical resonator}.  These expressions are then used in section \ref{sec:results_ana} for comparing the modeling results. In the same section, an empirical expression for tunneling distance of fundamental modes of a toroidal cavity is also provided. In section \ref{sec:results_exp}, the model is validated experimentally by measuring the quality factors of the microtoroidal cavities in liquid. In section \ref{sec:results_num}, the model is validated numerically by running convergence test. In section \ref{sec:disc}, we show that the model can be applied for sensing applications. In this section, we show that for a sensor, whose sensing metric is change in the quality factor, an optimum geometry exists for achieving maximum sensitivity.

\section{Mathematical description}
\label{sec:math}
Applying Galerkin's method to the wave equation and after using the boundary conditions for open resonators, one can arrive at the FEM equation in the weak form \cite{Oxborrow2007}:

\begin{equation}
\int _V \! \left  (\vec{\nabla} \times \vec{\widetilde H^*})\epsilon^{-1}(\vec{\nabla} \times \vec{H}) - \alpha(\vec{\nabla} \cdot \vec{\widetilde H^*})(\vec{\nabla} \cdot \vec{H})+ c^{-2}\vec{\widetilde H^*}\cdot \dfrac{\partial^2 \vec{H}}{\partial t^2} \right ) \, dV=0
\label{Eqn:FEM Equation}
\end{equation}
where $\vec{H}$ represents the magnetic field of the resonator and $\vec{\widetilde H}$ represents the test magnetic field, an essential component of the weak form. The second term of Eq. (\ref{Eqn:FEM Equation}) represents a penalty term to suppress false solutions. None of the field components will depend upon the azimuthal coordinate $\phi$ in the axisymmetric resonators, resulting in reduction of the 3D problem to a 2D problem.
\subsection{Perfectly matched layer formulation}
\label{sec:PML}
A PML can be treated as an anisotropic absorber  in which the diagonal permittivity and permeability tensors of the absorber are modified according to Eq. (\ref{Eqn:Perm. modification}) \cite{Chew1997}.

\begin{equation}
\bar{\epsilon}=\epsilon \bar{\Lambda}, \bar{\mu}=\mu \bar{\Lambda}
\label{Eqn:Perm. modification}
\end{equation}

The radial and axial modification factors are represented by $\bar {\Lambda}$, which is given by Eq. ( \ref{Eqn:Lambda}) 
\begin{equation}
\bar {\Lambda}=\left (\dfrac{\tilde r}{r} \right) \left (\dfrac{s_z}{s_r} \right )\hat{r}+ \left (\dfrac{r}{\tilde r} \right) (s_z s_r)\hat{\phi}+\left (\dfrac{\tilde r}{r} \right ) \left (\dfrac{s_r}{s_z} \right )\hat{z}
\label{Eqn:Lambda}
\end{equation}

where
\begin{equation}
s_r=
\begin{cases}
n_{medium} & \text{$0 \leq r \leq r_{pml}$}\\
n_{medium} -j G \left (\dfrac {r-r_{pml}}{t_{rpml}} \right )^N & \text{$r > r_{pml}$}
\end{cases}
\end{equation}
\begin{equation}
s_z=
\begin{cases}
n_{medium}-j G \left (\dfrac {z_{lpml}-z}{t_{lpml}} \right )^N  &\text{$z < z_{lpml}$}\\
n_{medium} &\text{$z_{lpml} \leq z \leq z_{upml}$}\\
n_{medium}-j G \left (\dfrac {z-z_{upml}}{t_{upml}} \right )^N &\text{$z > z_{upml}$}\\
\end{cases}
\end{equation}
\begin{equation}
\tilde r=
\begin{cases}
r &\text{$0 \leq r \leq r_{pml}$}\\
r-j G \left (\dfrac {(r-r_{pml})^{N+1}}{(N+1)t_{pml}^N} \right ) &\text{$r > r_{pml}$}
\end{cases}
\end{equation}

where $t_{rpml},$ $t_{upml}$, $t_{lpml}$  are the PML thicknesses in the radial, +z and -z directions respectively and $r_{pml}$, $z_{upml}$, $z_{lpml}$ are the locations of the start of PML in the radial, +z and -z directions respectively. $n_{medium}$ is refractive index of the medium, $N$ is order of the PML, and $G$ is a positive integer.

In the PML expressions ($s_r,s_z, \tilde r$), the imaginary component contributes to the attenuation of waves in the PML but at the same time, due to the discrete nature of the FEM mesh, a large imaginary component will introduce reflections at the interface between the PML and the medium. In order to determine the optimal value for the imaginary component, we have investigated linear, quadratic, and cubic PML of different thicknesses for various values of $G$ by running many simulations for various sphere diameters. To deduce the optimum values of the parameters, we then compared the simulation results for $Q_{WGM}$ of spherical cavities with the analytical ones. The simulation results show that a linear (i.e. $N=1$), and $\lambda/4$ thick PML with a $G$ value of 5 is optimum. We have also used these optimum values for the simulations of the microtoroidal cavities. The location of PML is also important to obtain accurate results. The radial PML should be greater than the tunneling distance (t) of the microcavity and the z PML should be greater than FWHM ($w_z$) of WGM along the z axis. After running series of simulations we find the following as optimum values (w.r.t. $(r,z)=(0,0)$, see Fig. \ref{fig:potfield}):

\begin{equation}
 r_{pml}\ge 6t
\end{equation}
\begin{equation}
 z_{(u,l)pml}\ge |5.5w_z|
\end{equation}

\subsection{Finite element method equation with perfectly matched layer}
In order to incorporate the PML,  we have reformulated Eq. (\ref{Eqn:FEM Equation}) in the following way:
\begin{equation}
\int_V \! \left ( (\vec{\nabla} \times \vec{\tilde H^*})\bar{\epsilon}^{-1}(\vec{\nabla} \times \vec{H}) - \alpha(\vec{\nabla} \cdot \vec{\tilde H^*})(\vec{\nabla} \cdot \vec{H})+ c^{-2}\vec{\tilde H^*}\cdot \bar{\mu}\cdot \dfrac{\partial^2 \vec{H}}{\partial t^2} \right ) \, dV=0
\label{Eqn:FEM master Equation}
\end{equation}
By casting Eq. (\ref{Eqn:FEM master Equation}) into the FEM software COMSOL, a full vectorial finite element model of a silica sphere in air can be obtained. By using the eigenvalue solver in COMSOL, resonant frequencies ($f_r$) of all the modes can easily be determined. Quality factor due to the WGM radiation can be calculated as \cite{Oxborrow2007}:
\begin{equation}
Q_{wgm}= \dfrac{\Re{(f_r)}}{2\Im{(f_r)}}
\label{Eqn:Q_wgm}
\end{equation}
\subsection{Analytical expressions of the spherical resonator}
\label{sec:Analytical expressions of the spherical resonator}
\subsubsection{Quality factor}
The quality factor due to WGM radiation losses in a spherical microcavity can be written as \cite {Datsyuk92}:
\begin{equation}
	Q_{wgm}=\dfrac{1}{2}(m+\dfrac{1}{2})p^{1-2M}(p^2-1)^{\frac{1}{2}}e^{2T_m} \mbox{,\hspace{5mm}  $m\gg 1$}
	\label{Eqn:Q_{wgm}}
\end{equation}
where\\
\begin{center}
    $m$ = azimuthal mode number
\end{center}
\begin{equation}
p^2 = \dfrac{\epsilon_{sphere}}{\epsilon_{medium}}
\end{equation}
\begin{equation}
T_m=(m+\dfrac{1}{2})(m_l-\tanh(m_l)
\end{equation}
\begin{equation}
m_l=\cosh^{-1} \left (p \left(1-\dfrac{1}{m+\frac{1}{2}}\left (t_q^0\beta +\dfrac{ p^{1-2M}}{\sqrt {(p^2-1)}} \right) \right )^{-1} \right )
\end{equation}
\begin{equation}
\beta =\left(\frac{1}{2}\left(m+\frac{1}{2}\right)\right)^{\frac{1}{3}}
\end{equation}
\begin{equation}
M=
\begin{cases}
0 &\text{For TE}\\
1 &\text{For TM}
\end{cases}
\end{equation}
$t_q^0$ is the $q^{th}$ root of equation $F_{airy}(t_q^0)=0$\\\\

$m$ can be calculated using the characteristic equation for WGM frequencies \cite{weinstein69}:
\begin{equation}
	p^{1-2M}\dfrac{j_m'[pk_0a]}{j_m[pk_0a]}=\dfrac{h_m'[k_0a]}{h_m[k_0a]}
	\label{Eqn:CharEqn}
\end{equation}
where $j$, $h$ are Bessel functions, $k_0$ is the wave number ($2\pi/\lambda_0$), and $a$ is the radius of a microsphere.

It should be noted that Eq. (\ref{Eqn:Q_{wgm}}) is an asymptotic solution for the $Q_{wgm}$  of the spherical resonator which requires $m\gg 1$, however the error is less than $1\%$ for $m \geq 19$  \cite{Buck2003}. In our comparison for the analytical results, we have applied Eq. (\ref{Eqn:Q_{wgm}}) to silica spheres with mode numbers (m) ranging from $25-80$.
\subsubsection{Tunneling distance}

The Schrodinger equation for a microsphere for a fundamental mode can be written in radial coordinates (r) as \cite{Johnson1993}:
\begin{equation}\label{eqn:schrondinger}
    -\dfrac{d^2 \Psi_r}{dr^2}+ V_r \Psi_r=E \Psi_r
\end{equation}
where $E=k^2$ is the total energy and $\Psi_r$ is a position probability function of a photon. The potential ($V_r$) is given by

\begin{equation}\label{eqn:pot}
    V_r=k^2(1-p_r^2)+\dfrac{m(m+1)}{r^2}
\end{equation}
where $p_r$ is ratio of relative permittivity of the sphere and the surrounding medium in radial coordinates. Analytically, tunneling distance for a fundamental mode of a microsphere can be written as \cite{Johnson1993}:
\begin{equation}\label{eqn:tun}
    t=\dfrac{\sqrt{m(m+1)}}{k}-a
\end{equation}

\section{Results}
\label{sec:results}
\subsection{Comparison between simulations and analytical results}
\label{sec:results_ana}
\begin{figure}[htp]

\centering
{
    \includegraphics[width=11cm]{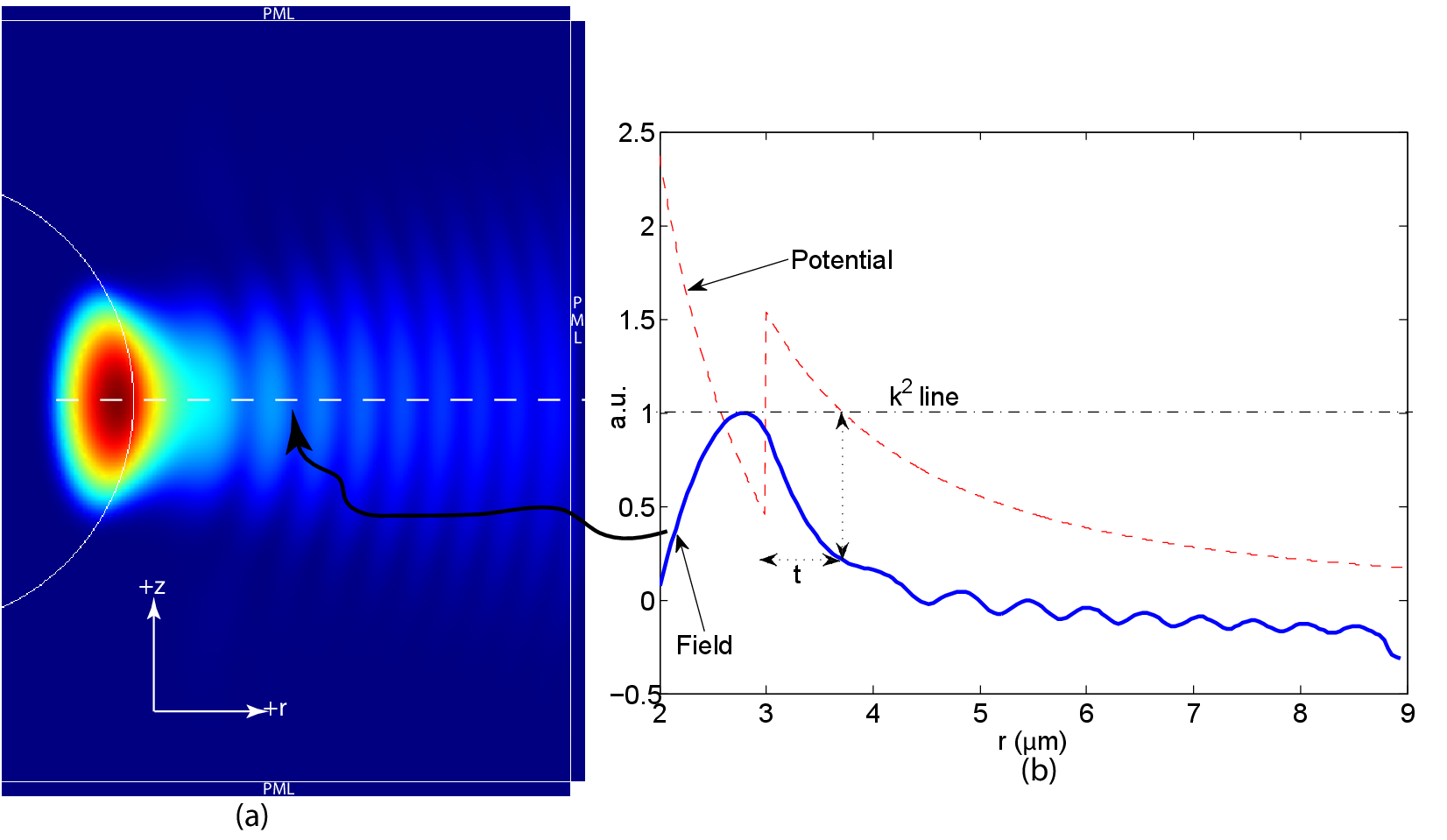}
}
\caption{a) (False color) Logarithmic intensity of the fundamental TE mode of a silica microsphere in air. b) Normalized field intensity along the dashed line shown in (a). Dotted curve shows potential, normalized to $k^2$. Tunneling distance is indicated by t.}
\label{fig:potfield}
\end{figure}

Figure \ref {fig:potfield} shows a fundamental TE mode of a silica spherical cavity in air.

\begin{figure}
\centering
{
    \label{}
    \includegraphics[width=6.25cm]{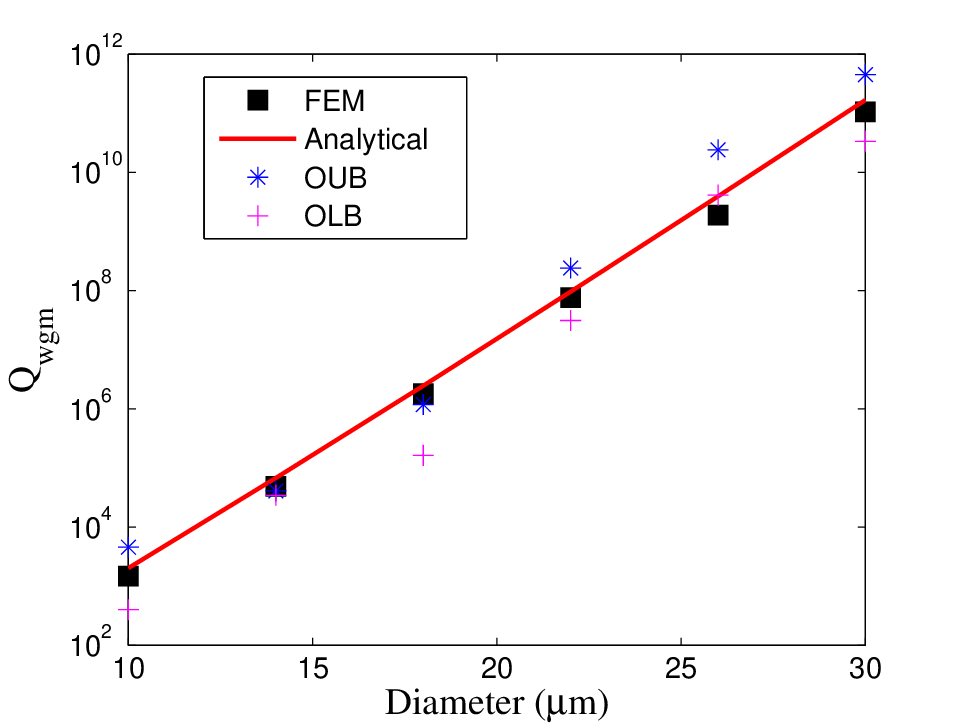}
}
\caption{Modeling results at $1550nm$. WGM Quality factors for various silica sphere diameters of fundamental TE mode. OUB and OLB represent upper and lower bounds that are calculated by using Oxborrow's model \cite{Oxborrow2007}.}
\label{fig:Q} 
\end{figure}

We have plotted the quality factor due to the fundamental TE whispering gallery mode radiation for various sphere diameters at $1550nm$. Similar results are obtained for the TM mode. Figure \ref{fig:Q} shows the comparison of the FEM simulation results and results obtained by using analytical expressions presented in section \ref{sec:Analytical expressions of the spherical resonator}. We have also calculated the minimum and maximum $Q_{wgm}$ values using Oxborrow's model \cite{Oxborrow2007} for each sphere diameter and those are also shown in Fig. \ref{fig:Q}. It can be seen that the new model provides a more accurate estimate than Oxborrow's model, and returns a single value of Q rather than a range of values. The slight difference between the analytical and the FEM values is attributed to discretization of the computation domain. The accuracy will improve with finer mesh elements (see section \ref{sec:results_num}). It can also be seen that Oxborrow's bounds do not always straddle the analytical solutions. One of the possible reasons for this discrepancy may be that while deriving these bounds, in order to simplify the derivation,  Oxborrow assumes that the modes are transverse but in reality the modes of these axisymmetric resonators are not perfectly transverse \cite{Oxborrow2007}.

\begin{figure}
\centering
\subfigure[Microsphere]
{
    \label{fig:sphere_t}
    \includegraphics[width=6.25cm]{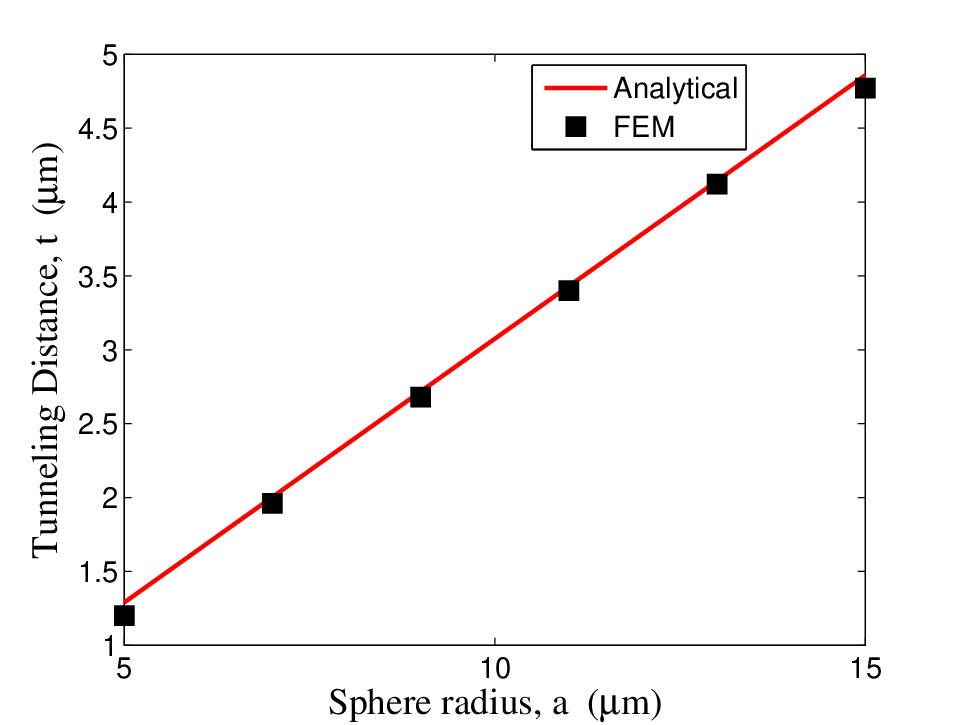}
}
\hspace{0.1cm}
\subfigure[Microtoroid]
{
    \label{fig:tor_t}
    \includegraphics[width=6.25cm]{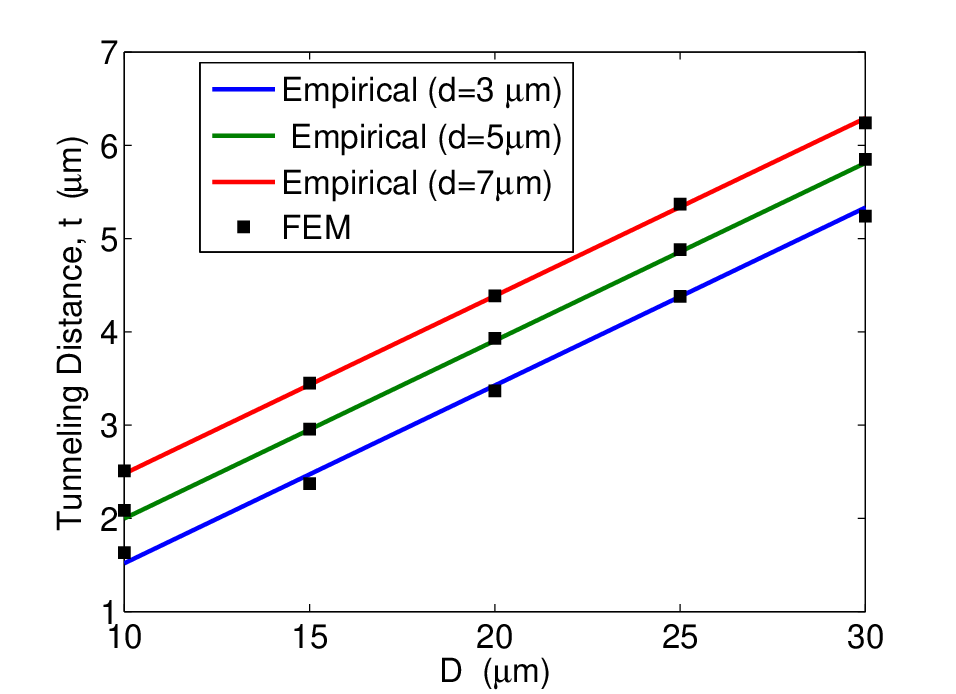}
}
\caption{Modeling results at $1550nm$. Tunneling distance of a fundamental TE mode as a function of the microcavity geometries. Inset of Fig. \ref {fig:tor_t} shows the cross section of a microtoroidal cavity. }
\label{fig:t} 
\end{figure}

Figure \ref{fig:t} shows the results for tunneling distance for both microsphere and microtoroidal cavities. An empirical relation  for the tunneling distance for fundamental modes of the microtoroidal cavities can be extracted from the simulation results and is given as:
\begin{equation}\label{eqn:micro_tun}
    t=\dfrac{\sqrt{m(m+1)}}{k} - \dfrac{D+d}{2}
\end{equation}
where $D$ and $d$ are major and minor diameter of a microtoroid and are shown in Fig. \ref {fig:tor_t}. The tunneling distance results are also in agreement with experimental results presented in \cite{Carmon2009}.

\subsection{Comparison between simulations and experimental results}
\label{sec:results_exp}
Our model is also appropriate for other axisymmetric resonators where analytical solutions are not available. In order to test this, we have used our FEM model to determine the total quality factor of microtoroidal cavities \cite{Armani2003} and have compared the simulation results with experimental measurement of the quality factor. Mathematically the total quality factor ($Q_{total}$) can be represented by:

\begin{equation}
	 Q_{total}=\dfrac{1}{Q_{wgm}^{-1}+Q_{surroundings}^{-1}+Q_{material}^{-1}+Q_{coupling}^{-1}}
	\label{Eqn:Q_{total}}
\end{equation}

 In order to validate the model with reasonable range of quality factors, it is necessary to ensure suitable experimental conditions in which to observe these quality factors. We have performed the experiments with the microtoroids immersed in ethanol (refractive index: $1.3538$ at $1550nm$ \cite{Emmerson2009}) by using the fluidic cell described in \cite{kirk2012}.  We have preferred ethanol over water as ethanol not only provides low refractive index contrast between the silica cavity and the surroundings but also has low absorption  as compared to water at $1550nm$ \cite{Guardia1993}. This will allow us to select a wide range of microtoroidal cavities whose $Q_{total}$ is mainly limited by the $Q_{wgm}$.

\begin{figure}[htp]
\centering
{
    \includegraphics[width=6.25cm]{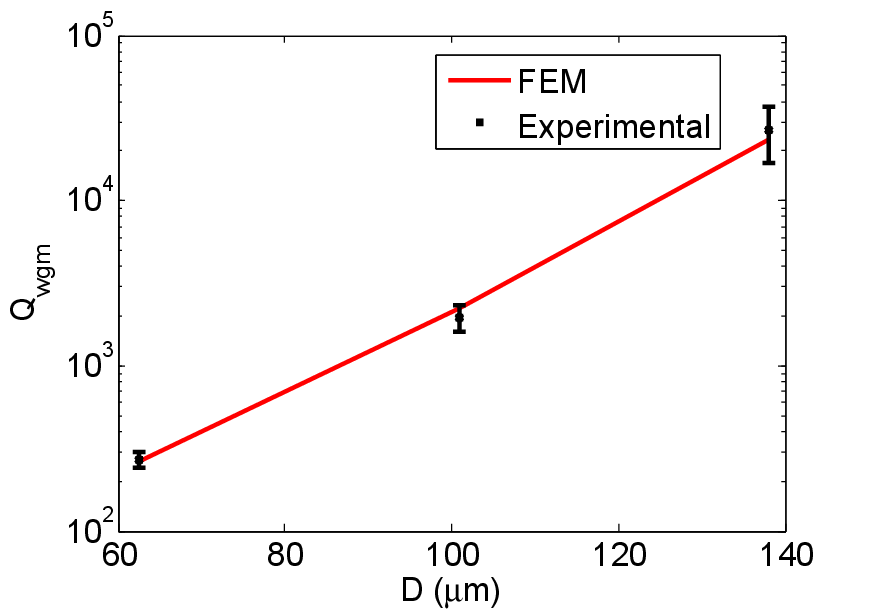}
}
\caption{Comparison between the modeling and the experimental results. WGM Quality factors of fundamental TM mode for various silica microtoroidal cavity diameters immersed in ethanol. Minor diameter $d$ of each cavity is slightly different and is around $5\mu m\pm 1\mu m$. The experimental quality factors are determined by Lorentz curve fitting of the resonant peaks ($Q=\lambda/\Delta \lambda$) where $\lambda\approx1530nm.$}
\label{fig:TM_exp}
\end{figure}

 A broadband source (peak wavelength:$1530nm$) is used to couple the light into microtoroids via a tapered optical fiber (taper waist: $\le 1\mu m$ ). The resonant peaks are observed on a high resolution optical spectrum analyzer (Apex AP 2443B, resolution: $0.16pm$ ) and quality factor is determined by Lorentz curve fitting of the peaks ($Q=\lambda/\Delta \lambda$). The tapered fiber is positioned (using $10nm$ resolution nanostages) along the equator of the cavity to couple light into fundamental transverse modes. It is ensured that the tapered fiber does not touch the cavity during the measurements.  In order to minimize the effect of $Q_{coupling}$  on $Q_{total}$, the power at the peak wavelength of the source is set around $-48dbm$ (Optical spectrum analyzer sensitivity: $-70dbm$) and all the measurements are taken in highly undercoupled regime. The comparison between simulation and experimental results is shown in Fig. \ref{fig:TM_exp}.

\subsection{Computational speed and numerical accuracy}
\label{sec:results_num}
For the results presented in Figs.\ref{fig:potfield}-\ref{fig:TM_exp}, we have used the quadratic Lagrange elements of triangular shape with average element size (longest edge) of $0.3 \mu m$ and the number of degrees of freedom of order $10^5$.  As an example, a microtoroidal cavity model (Computation domain size:$25 \mu m \times 16 \mu m$, $D=30\mu m, d=5 \mu m$) with the aforementioned statistics took only $35s$ to find the first $20$ eigenvalues on a quad core $64$ bit operating system PC.

\begin{figure}[htp]

\centering
{
    \includegraphics[width=6.25cm]{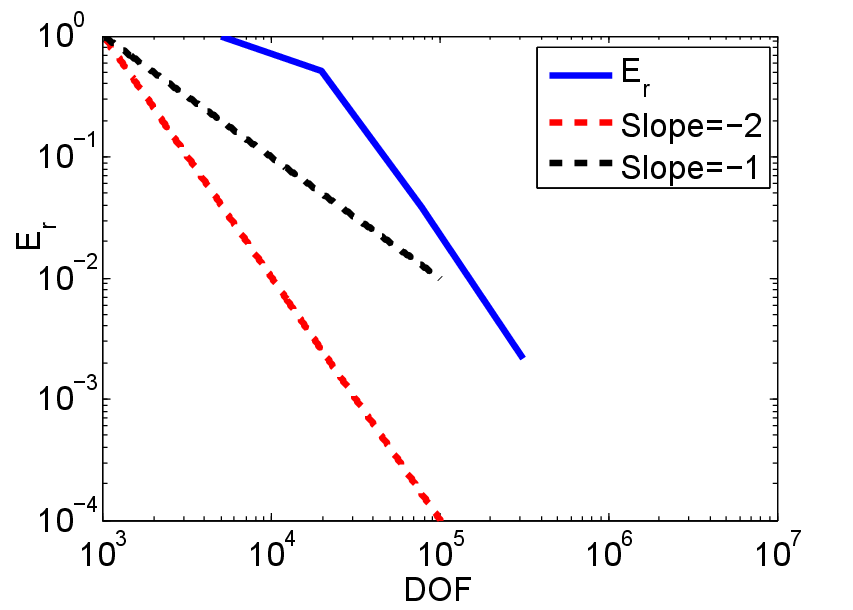}
}
\caption{Convergence plot for the quality factor determination of a microsphere.}
\label{fig:convergence}
\end{figure}

To check numerical accuracy of the model, we have run a convergence test. Figure \ref{fig:convergence} shows the plot of relative error ($E_r=(Q_{FEM}-Q_{exact})/Q_{exact}$)  as a function of the degrees of freedom (DOF) of our model. The DOF is related to the number of mesh elements, and the basis functions used in the FEM solver. Convergence of the solution is of order 2 which is expected for the Galerkin method for quadratic Lagrange elements \cite{Valli1997}. It can be seen in the Fig. \ref{fig:convergence} that for small DOF, convergence rate is slower than the expected; the reason for this is the poor approximation of the curved boundary of the cavity by the coarsely meshed triangular elements.

\section{Discussion and application to PS-CRDS microcavity sensor}
\label{sec:disc}
The results presented in section 3 show that our finite element model is both physically and numerically accurate. The quality factors for all the modes (fundamental and higher order modes) are also obtained by one single simulation rather than multiple simulations, which was the case for the original Oxborrow model. Moreover, no prior knowledge of any of the mode frequencies is required to obtain the quality factors. The model also gives fast results and without any fitting algorithms.

It should be noted that we have neglected dispersion for estimating the quality factors and tunneling distances. Whilst dispersion must be considered for very high Q applications where an equidistant modal spectrum is required \cite{Ilchenko_2003}, it has negligible effect for biosensing applications.

\begin{figure}[ht]
\centering
\subfigure[]
{
    \label{fig:op_pa}
    \includegraphics[width=5.5cm]{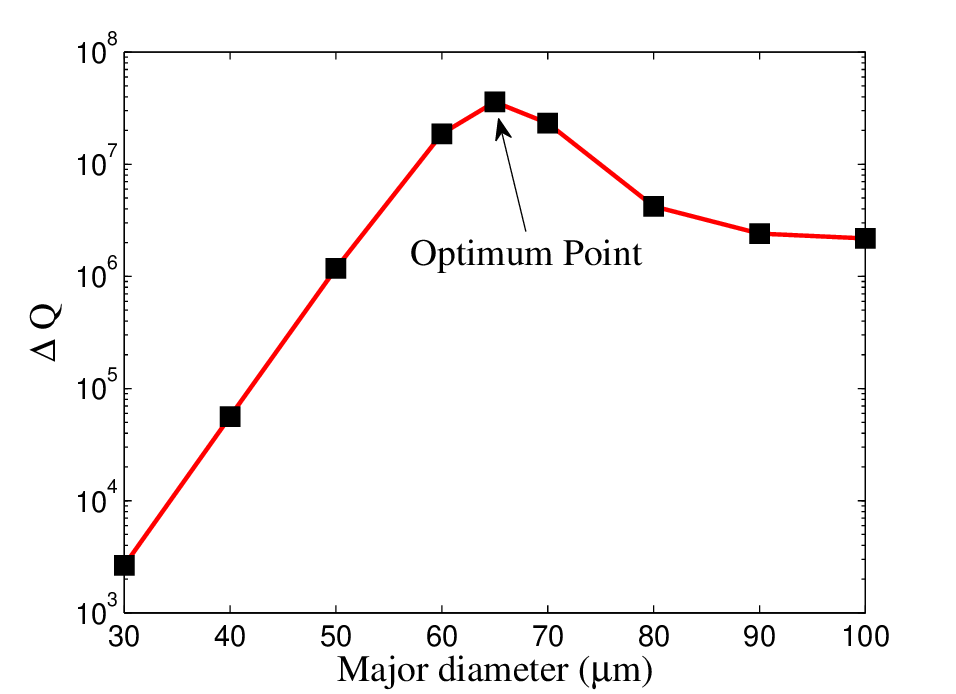}
}
\hspace{0.1cm}
\subfigure[]
{
    \label{fig:op_pb}
    \includegraphics[width=5.5cm]{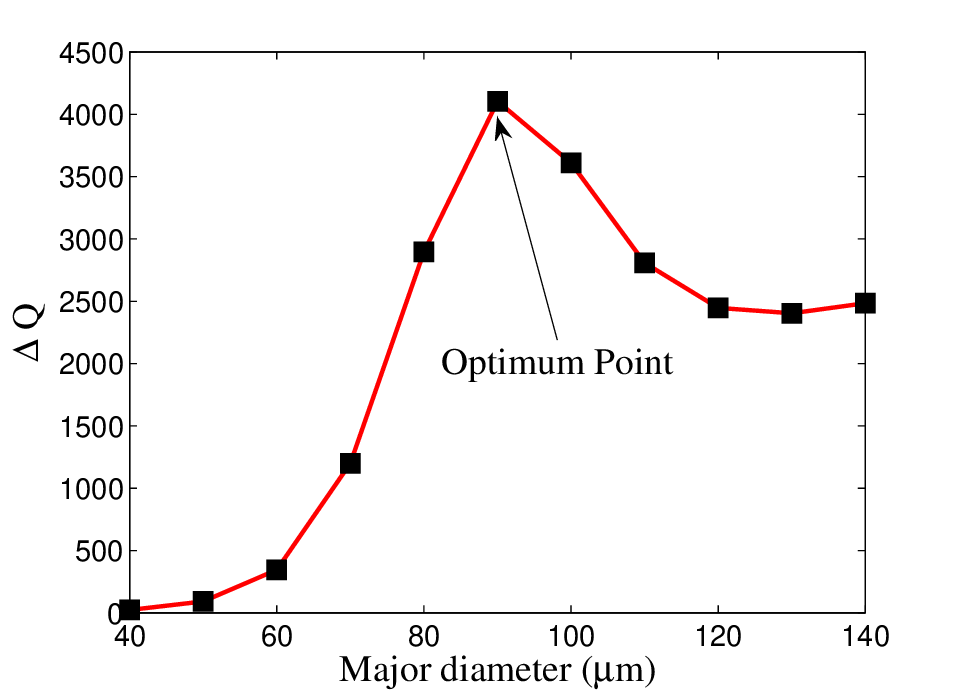}
}
\caption{Modeling results show the existence of an optimum geometry . Change in quality factor ($\Delta Q$) of fundamental TM mode as a function of major diameter (D) of a microtoroidal cavity (minor diameter$(d)=6\mu m$) immersed in water. $\Delta Q$ shows the difference between $Q_{total}$ for water and that measured when small refractive index change ($\delta n=10^{-3}$) is introduced into the water. (a)Modeling results for $\lambda=633nm$ (b)Modeling results for $\lambda=1530nm$.}
\label{fig:op_p}
\end{figure}

We have also applied our model to the PS-CRDS microtoroidal cavity sensor \cite{kirk2012} in order to determine the optimum parameters for an application of refractometric sensing. In such a scheme, the microcavity is immersed into water and a small refractive index change ($\delta n$) is introduced. This change in refractive index will influence the $Q_{total}$ of the cavity and can easily be measured via PS-CRDS microtoroidal cavity sensor \cite{kirk2012}. Figure \ref{fig:op_p} shows the modeling results for change in $Q_{total}$ of the cavity as a function of $D$. It can be clearly seen that to achieve maximum sensitivity there exists an optimum geometry for both $\lambda=633nm$ and $\lambda=1530nm$. Due to high water absorption at $1530nm$, $\Delta Q$ is lower than the one at $633nm$.  Since $Q_{total}$ varies with the wavelength, the optimum geometry will be different for each of the wavelength.

\begin{figure}[ht]
\centering
\subfigure[]
{
    \label{fig:qt_a}
    \includegraphics[width=5.5cm]{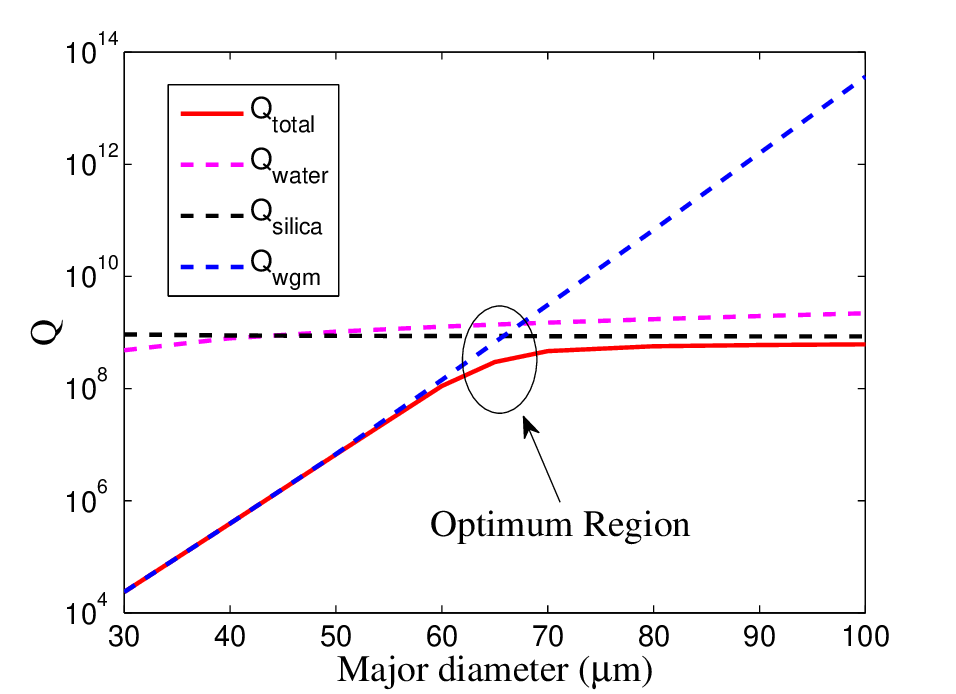}
}
\hspace{0.01cm}
\subfigure[]
{
    \label{fig:qt_b}
    \includegraphics[width=5.5cm]{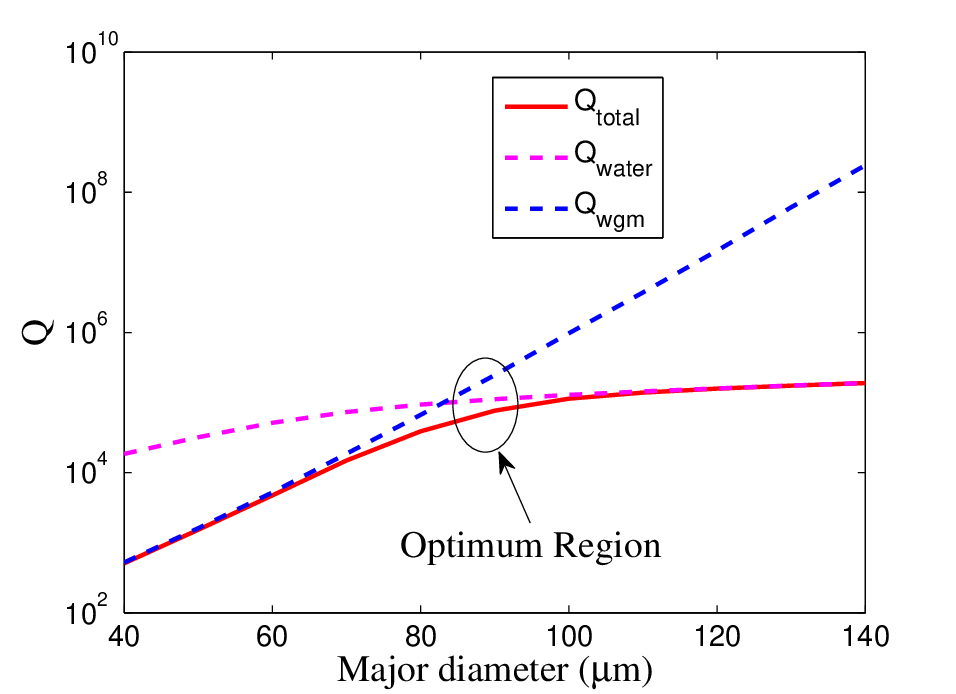}
}
\caption{Modeling results.  Various quality factors (Eq. (\ref{Eqn:Q_{total}})) for fundamental TM mode involved in a microtoroidal cavity immersed in water as a function of $D$ ($d=6\mu m$). In the optimum region the $Q_{wgm}$ is close to $Q_{total}$.(a) Modeling results at $\lambda=633nm$. (b) Modeling results at $\lambda=1530nm$. $Q_{silica} \approx 10^{11}$ (Silica has very low absorption at $1530nm$) and is omitted in the figure to reduce the range in y-axis.}
\label{fig:Qt}
\end{figure}

In order to understand that why an optimum geometry exists for a $\Delta Q$ measurement, we have plotted the individual terms of Eq.  (\ref{Eqn:Q_{total}}) as a function of the cavity geometry (Fig. \ref {fig:Qt}). We have assumed $Q_{coupling}=0$ which is a reasonable assumption as experimentally, contribution of $Q_{coupling}$ can be minimized by taking measurements at multiple input power levels \cite{Armani2005c}. Figure \ref {fig:Qt} shows that the optimum region exists close to the point of inflection of the $Q_{total}$ curve. Results in Fig. \ref{fig:Qt} show that as diameter increases, the  WGM is better confined within the microcavity (see Fig. \ref{fig:potfield}(a)) and so $Q_{wgm}$ is less influenced by the external environment, with the result that the total quality factor is limited by silica absorption.

Currently the standard experimental approach to determine the total quality factor of a microcavity  is either by  cavity ring down spectroscopy or  Lorentzian fitting of a resonant peak. However, tunneling distance is also related to WGM radiative quality factor i.e. large tunneling distance means high $Q_{wgm}$ and vice versa. This suggests that  with model provided here coupled with the experimental procedure outlined in \cite{Carmon2009}, one can also extract $Q_{wgm}$ even when $Q_{total}$ is greater than $10^8$ . Such a measurement can possibly open up the way to a new sensing method for microcavity sensors.

We have formulated an empirical expression (Eq. (\ref{eqn:micro_tun})) for the tunneling distance of the fundamental modes of microtoroids. This expression is inspired by the analytical expression for the tunneling distances of microspheres by treating the microtoroid as a sphere of diameter ($D+d$). This formulation is not surprising as these equations are true only for fundamental modes which lie along the equatorial region of a microcavity. However, this treatment will not be true for higher order modes (non equatorial modes) as the curvature for a microsphere of diameter $D+d$ will be quite different from a microtoroid with $D$ and $d$ as its dimensions (see Fig.\ref{fig:t}(b)).

In summary, our finite element model, without any approximation in its master equation and coupled with a PML, not only gives accurate quality factors and tunneling distances, but can also determine the other important parameters (e.g. mode volumes) accurately for a wide range of applications based on axisymmetric microcavities such as biosensing, non-linear processes, and lasers.

\section*{Acknowledgments}
We thank Prof. Jonathan Webb at McGill University for useful discussions which assisted in our understanding of PMLs. We are grateful to Prof. Andrea Armani at University of Southern California for her invaluable suggestions for improving the manuscript.  We appreciate the efforts of Ashley Maker at University of Southern California, for fabricating the microtoroidal cavities used in this work. We are also grateful to Prof. Tal Carmon at University of Michigan for pointing us to tunneling distance calculations.  This work is supported by the NSERC-CREATE training program in Integrated Sensor Systems, McGill Institute of Advanced Materials, Montreal, Canada.


\begin{thebibliography}{1}
\newcommand{\enquote}[1]{``#1''}
\bibitem{Vollmer2012} F. Vollmer and L. Yang, ``Label-free detection with high-Q microcavities: a review of biosensing mechanisms for integrated devices,'' Nanophotonics 267--291 (2012).

\bibitem{Fan_2008}
X. Fan, I. M. White, S. I. Shopoua, H. Zhu, J. D. Suter, and Y. Sun,``Sensitive optical biosensors for unlabeled targets: A review,'' {{Analytica Chimica Acta}} \textbf{620}, 8--26 (2008)


\bibitem{Ilchenko_P2002}

 L. Maleki and  V. S. Ilchenko, ``Techniques and devices for sensing a sample by using a whispering gallery mode resonator,''  California Institute of Technology, US Patent 6,490,039, (2002).

\bibitem{Ilchenko_2002}
J. L. Nadeau, V. S. Ilchenko, D. Kossakovski, G. H. Bearman, and L. Maleki, \enquote{High-Q whispering-gallery mode sensor in liquids,} Proc. SPIE {\bf4629,} 72   ( {2002}).

\bibitem{Armani_2006}
A. M. Armani and K. J. Vahala, \enquote{Heavy water detection using ultra-high-q microcavities,} Opt. Lett. \textbf{31}, 1896--1898 (2006).

\bibitem{Barnes_2008}
J. Barnes, B. Carver, J. M. Fraser, G. Gagliardi, H. P. Loock, Z. Tian, M. W. B. Wilson, S. Yam, and O. Yastrubshak, \enquote{Loss determination in microsphere resonators by phase-shift cavity ring-down measurements,} {Opt. Express} \textbf{{16}}, {13158--13167} ({2008}).

\bibitem{kirk2012} M. I. Cheema, S. Mehrabani, A. A. Hayat, Y. A. Peter, A. M. Armani and A. G. Kirk, ``Simultaneous measurement of quality factor and wavelength shift by phase shift microcavity ring down spectroscopy,'' Opt. Express \textbf{20}, 9090-9098 (2012).


\bibitem{Spielmann1994}C. Spielmann, R. Szip\"ocs,  A. Stingl, and F. Krausz,  ``Tunneling of optical pulses through photonic band gaps,'' Phys. Rev. Lett. \textbf{73}, 2308--2311, (1994).

\bibitem{Pedro2000}  P. Pereyra, ``Closed formulas for tunneling time in superlattices,'' Phys. Rev. Lett. \textbf{84}, 1772--1775 (2000).

\bibitem{Podolskiy2005}V. A. Podolskiy, and E. E. Narimanov, ``Chaos-assisted tunneling in dielectric microcavities,'' Opt. Lett. \textbf{30}, (2005).

\bibitem{Carmon2009}M. Tomes, K. J. Vahala, and T. Carmon, ``Direct imaging of tunneling from a potential well,'' Opt. Express  \textbf{17}, 19160 (2009).

\bibitem{Chinellato2005} O. Chinellato, P. Arbenz, M. Streiff, and A. Witzig, ``Computation of optical modes in axisymmetric open cavity resonators,'' Future Gener. Comp. Sy. \textbf{21}, 1263--1274 (2005).

\bibitem{Karl2009} M. Karl, B. Kettner, S. Burger, F. Schmidt, H. Kalt, and M. Hetterich, ``Dependencies of micro-pillar cavity quality factors calculated with finite element methods,'' Opt. Express \textbf{2}, (2009).

\bibitem{Suzuki1985} M. Koshiba,  K. Hayata, and M. Suzuki, ``Improved finite element formulation in terms of the magnetic field vector for dielectric waveguides,'' IEEE Trans. on Microwave Theory Tech. \textbf{33}, 227-233, (1985).

\bibitem{Osegueda94} R. A. Osegueda, J. H. Pierluissi, L. M. Gil, A. Revilla, G. J. Villalva, G. J. Dick, D. G. Santiago, and R. T. Wang, ``Azimuthally dependent finite element solution to the cylindrical resonator,''10th annual review of progress in applied computational electromagnetics \textbf{ 1}, 159--170, (1994).

\bibitem{Oxborrow2007}  M. Oxborrow, ``Traceable 2-D finite element simulation of the whispering gallery modes of axisymmetric electromagnetic resonators,'' IEEE Trans. on Microwave Theory Tech. \textbf{55}, 1209--1218 (2007).

\bibitem{Spillane2004}S. M. Spillane, ``Fiber-coupled ultra-high-Q microresonators for nonlinear and quantum optics,'' PhD. thesis, CALTECH (2004)

\bibitem{Soltani2010}M. Soltani, Q. Li, S. Yegnanarayanan, and A. Adibi,``Toward ultimate miniaturization of high Q silicon traveling-wave microresonators,'' Opt. Express \textbf{18}, 19541-19557 (2010).

\bibitem{Oxborrow2012}\text{https://sites.google.com/site/axisymmetricmarkoxborrow/home}

\bibitem{Oxborrow2012a}M. Oxborrow, J. D. Breeze and N. M. Alford, ``Room-temperature solid-state maser,'' Nature \textbf{488}, 353–356 (2012).

\bibitem{Armani2012} C. Shi, H. S. Choi, A. M. Armani,``Optical microcavities with a thiol-functionalized gold nanoparticle polymer thin film coating,'' Appl. Phys. Lett. \textbf{100}, 013305 (2012).

\bibitem{Vahala2011} T. Lu,  H. Lee, T. Chen, S. Herchak, J. H.  Kim, S. E. Fraser, R. C. Flagan, and K. Vahala, ``High sensitivity nanoparticle detection using optical microcavities,'' Proc. Natl. Acad. Sci. (2011).


\bibitem{Gong2010}Y. F. Xiao, C. L. Zou, B. B. Li, Y. Li, C. H. Dong, Z. F. Han, and Q. Gong, ``High-Q exterior whispering-gallery modes in a metal-coated microresonator,'' Phys. Rev. Lett. \textbf{105}, 153902, (2010).

\bibitem{Kirk2010}M. I. Cheema, and A. G. Kirk. ``Implementation of the perfectly matched layer to determine the quality factor of axisymmetric resonators in COMSOL,'' in {\it COMSOL conference,} Boston, Oct 8 2010.

\bibitem{Goddard2010} A. Arbabi, Y. M. Kang, and L. L. Goddard, ``Analysis and Design of a Microring Inline Single Wavelength Reflector,'' in {\it FIO,} October 24, 2010.

\bibitem{Chew1997} F. L. Teixeira, and W. C. Chew, ``Systematic derivation of anisotropic PML absorbing media in cylindrical and spherical coordinates,'' IEEE Microwave Guided Wave Lett. \textbf{7}, 371-373, (1997).

\bibitem{Datsyuk92} V. V. Datsyuk, ``Some characteristics of resonant electromagnetic modes in a dielectric sphere,''Appl. Phys. \textbf{B54,} 184--187, (1992).

\bibitem{weinstein69} L. A. Weinstein, {\it Open Resonators and Open Waveguides} (The Golem Press, Boulder, Colorado, USA, 1969) pp. 298.

\bibitem{Buck2003}J. R. Buck, and H. J. Kimble, `` Optimal sizes of dielectric microspheres for cavity QED with strong coupling,'' Phys. Rev. A \textbf{67} 033806 (2003).

    \bibitem{Johnson1993} B. R. Johnson, ``Theory of morphology-dependent resonances: shape resonances and width formulas,''J. Opt. Soc. Am. A \textbf{10}, 2, (1993).

\bibitem{Armani2003} D. K. Armani, T. J. Kippenberg, S. M. Spillane, and K. J. Vahala, ``Ultra-high-Q toroid microcavity on a chip,'' Nature \textbf{421}, 925--928 (2003).

\bibitem{Emmerson2009}S. Scheurich, S. Belle, R. Hellmann, S. So, I.J.G. Sparrow, G. Emmerson, ``Application of a silica-on-silicon planar optical waveguide Bragg grating sensor for organic liquid compound detection,'' Proc. SPIE \textbf{7356} (2009)

\bibitem{Guardia1993}M. Gallignani, S. Garrigues, and M. Guardia,``Direct determination of ethanol in all types of alcoholic beverages by near-infrared derivative spectrometry,''Analyst \textbf{118}, (1993).

\bibitem{Valli1997} A. Quarteroni, and A. Valli,  {\it Numerical Approximation of Partial Differential Equations}  (Springer, 1997).

\bibitem{Ilchenko_2003} V. S. Ilchenko, A. A. Savchenkov, A. B. Matsko, and L. Maleki, ``Dispersion compensation in whispering-gallery modes,'' J. Opt. Soc. Am. A  \textbf{20}, 157-162, (2003).

\bibitem{Armani2005c}A. M. Armani, D. K. Armani, B. Min, K. J. Vahala, and S. M. Spillane, ``Ultra-high-Q microcavity operation in \text{$H_2O$} and \text{$D_2O$}," Appl. Phys. Lett. \textbf{87}, (2005).
\end{thebibliography}
\end{document}